# Demonstrating Linked Battery Data To Accelerate Knowledge Flow in Battery Science

Philipp Dechent[z,1], Elias Barbers[2], Simon Clark[3], Susanne Lehner[4], Brady Planden[1], Masaki Adachi[1], David A. Howey[1,6], Sabine Paarmann[5]

[1] Department of Engineering Science, University of Oxford, OX1 3PJ, Oxford, UK

[2] Helmholtz Institute Münster (HI MS), IMD-4, Forschungszentrum Jülich, Campus-Boulevard 89, 52066 Aachen, Germany

[3] Battery Technology, SINTEF Industry, Bratsbergvegen 5, Trondheim 7031, Norway

[4] Department Elektrotechnik-Elektronik-Informationstechnik (EEI), Friedrich-Alexander-Universität Erlangen-Nürnberg, Cauerstraße 9, 91058, Erlangen, Germany

[5] Department of Mechanical Engineering, Imperial College London, UK.

[6] Faraday Institution, Quad One, Harwell Campus, Becquerel Ave, Didcot OX11 0RA, UK.

[z] Corresponding author philipp.dechent@gmail.com

**Abstract**

Batteries are pivotal for transitioning to a climate-friendly future, leading to a surge in battery research. Scopus (Elsevier) lists 14,388 papers that mention "lithium-ion battery" in 2023 alone, making it infeasible for individuals to keep up. This paper discusses strategies based on structured, semantic, and linked data to manage this information overload. Structured data follows a predefined, machine-readable format; semantic data includes metadata for context; linked data references other semantic data, forming a web of interconnected information. We use a battery-related ontology, BattINFO to standardise terms and enable automated data extraction and analysis. Our methodology integrates full-text search and machine-readable data, enhancing data retrieval and battery testing. We aim to unify commercial cell information and develop tools for the battery community such as manufacturer-independent cycling procedure descriptions and external memory for Large Language Models. Although only a first step, this approach significantly accelerates battery research and digitalizes battery testing, inviting community participation for continuous improvement. We provide the structured data and the tools to access them as open source.

**Highlights:**

- Large growth in battery research: In 2023, over 14,000 papers mentioning "lithium-ion battery" were published, highlighting the need for new data management and knowledge representation approaches.
- Introduction of structured and semantic data: we present a methodology using structured, semantic, and linked data to organise and access battery research efficiently.
- Enhanced data retrieval and automation: By developing machine-readable data sheets and employing full-text search capabilities, our approach accelerates battery research and moves towards automating battery testing.

1. **Introduction**

Batteries are one piece of the puzzle that must be solved for the transition to climate neutrality and a future without $CO_2$ emissions. Consequently, the battery research community is growing, and battery technology is developing rapidly. The number of research papers published each year is increasing; the number of articles containing the keyword phrase "lithium-ion battery" on the Scopus database [1] from Elsevier was 14,388 for the year 2023—on average that is around 40 papers per day, including weekends. At this rate, it is impossible for researchers to manage and comprehend each article or even simply judge if an article is relevant for them. This emphasises the need for alternative approaches to assess battery content.

In other fields, such as bioinformatics, standardised structured data with ontologies (i.e., formal frameworks to ensure consistent meaning) have been established [2]. These provide examples of how structured data helps comprehension of large amounts of complex information. The gene ontology, developed in 2000, established a common vocabulary accessible for humans and machines [2]. It is used as the foundation for a comprehensive database that links genes and human diseases. This database structure enables easy use for a broad audience and is compatible with various analysis tools [3].

The concept of structured, semantic and linked data is a valuable approach for other fields, such as batteries. This is closely related to the FAIR principles (Findability, Accessibility, Interoperability, and Reusability) which are designed to support knowledge discovery and innovation. They provide guidelines on data management with data being very broadly defined as any research output [4]. Adhering to these principles is much more challenging without a standard agreed on within the battery research community. Although here we create mainly linked data, we will use the term structured data in this paper more generally. More formally, we offer the following definitions:

(i) *Structured data* follows a predefined schema and is stored in a machine-readable format, e.g. an excel sheet, a parquet file, a JSON file.

(ii) *Semantic data* is structured data that includes metadata providing context and meaning. This data uses a shared and machine-readable vocabulary.

(iii) *Linked data* is semantic data that refers to other semantic data and creates a web of interrelated data. An example would be a JSON-LD description of a cell pointing to a JSON-LD description of the test data.

As the field of battery informatics is still young compared to other fields such as bioinformatics, the community lacks such unified procedures. This leaves the individual researcher with the challenge of identifying and filtering an overwhelming amount of data for relevant information. To address this and offer effective solutions, automation is necessary, propelled by advances in machine learning, large language models and structured battery data. The latter requires a clear set of vocabulary and corresponding dictionary to "understand" the text of a specific field. There are different relevant standards and recommendations like the International Electrotechnical Commission IEC 60050, IUPAC Gold Book, IEEE Battery Terminology, etc.; however, these standards are focused on facilitating human understanding with plain text descriptions. They do not provide the digital infrastructure necessary to integrate data into the Semantic Web, such as the Google Knowledge Graph and Wikidata. Therefore, work is in progress to develop a battery-related ontology. The Battery Interface Ontology (BattINFO) is a semantic resource with essential terms and relationships to describe electrochemical systems, materials, methods, and data. As an example, Figure 1 shows the entry for 'RatedCapacity' of a battery where the capacity value is declared by the manufacturer. The entry includes a unique identifier link and an elucidation that defines the term. Furthermore, if available, a reference to an existing standard is provided. Finally, additional details can be provided in a general parsable comment.

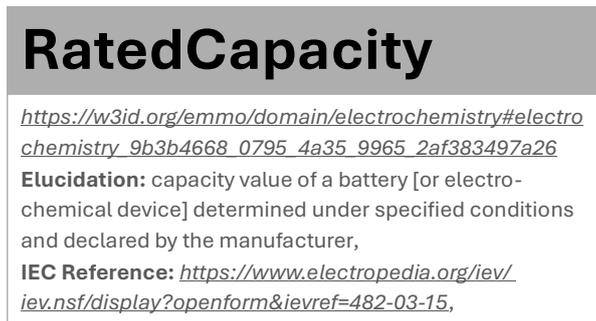

Figure 1: Data ontologies define consistent, standard frameworks for data, for example this entry for "RatedCapacity" in BattINFO. It also creates a node within the EMMO universe and in knowledge graphs. For example, machines know it is a physical quantity, with a given dimensionality, and accepts a certain set of units.

It must be admitted that defining "RatedCapacity" for members of the battery community might look like taking a step back, but it is immensely important to define a common language and understanding. Once this is achieved, the subsequent forward steps can be highly accelerated and automated.

To identify relevant papers from the flood of publications, databases (e.g. Scopus, CrossRef, Google Scholar, Semantic Scholar) offer search engines that scan the title, abstract and keywords for the search term. However, this does not enable a specific search for, e.g., the tested cells or cell chemistry, as this information is usually only stated in the experimental section. Using a text mining approach, El-Bousiydy et al. [5] estimated that using only the abstract for search purposes recovers only 11% of the possible information compared to the full text search of the same data set. Likewise, Westergaard et al. [6] found that the text mining of full texts was consistently better than using only abstracts by analysing 15 million publications in the field of biomedicine. Thus, a full text search is desirable. Alongside text mining from full texts, specific domain knowledge is required for meaningful data extraction. Consequently, battery-specific text mining tools have been developed [7]. These tools can be used to make information from literature available for computational and simulation applications.

Further, large language models (LLMs) can ease interaction with vast amounts of data and knowledge by using natural language to approximately retrieve the desired information. They are powerful and promising tools for accelerating research [8], however, the alignment of human reasoning and knowledge and model understanding and processing of information is challenging [9] and the tendency of LLMs to hallucinate [10] must be managed. Retrieval-augmented generation (RAG) is a promising approach as it provides the LLM with specific technical knowledge [11] and helps prevent it from hallucinating wrong key facts. RAG combines the parametric memory (e.g. neural model weights) of an LLM generated during the training process with external information (sometimes called 'non-parametric memory'). This approach uses the capability of the LLM to "understand" natural language and its general, basic knowledge. However, instead of relying on the knowledge gained during training, the LLM is equipped with additional resources to deal with knowledge-intensive queries [12]. In this work, we do not focus on the development of LLMs, but instead use available tools and develop the data sources to augment LLMs with specific battery knowledge organised as structured data. Although the rapid progress in the development of LLMs is exciting, it poses a further challenge to keep up to date. Thus, the use of structured data ensures that the knowledge contained can be used independently by new generations of LLMs.

The overall goal of this work is to advance battery research by implementing structured data methodologies that align with the FAIR principles—making data more Findable, Accessible, Interoperable, and Reusable. Therefore, this work presents a methodology to create structured data which enables us to deal with the information from thousands of papers per year and offers a possibility to make battery testing more digital and automated.

- Firstly, we enhance findability through semantic searches and linking publications and cell types. For this purpose, we use an exhaustive full text search of more than 80,000 articles to create structured data for batteries, comprising data files with information on commercial batteries which are additionally organised in a knowledge graph database for visualisation and comparison.
- Secondly, we demonstrate how the structured data can be complemented and accessed with the help of large language models and thus improves accessibility and reusability via retrieval-augmented generation for non-experts.
- Finally, this work promotes interoperability by introducing a Battery Cycler Language.

These advances are demonstrated in Case Studies and the resources and several examples are provided open source. So far, we started the process of standardisation focusing on lithium-ion batteries, but the methodology is transferable to other chemistries and battery types. Thus, this work propels the battery field toward achieving FAIR data standards.

## 2. Methods and implementation

The following section explains the method of creating a structured dataset that is machine- and human-readable to improve findability and reusability of the data. The focus here is on data and information about specific cells and experimental procedures. Therefore, we first introduce the ontology and semantic data treatment as the basis for a common understanding and introduce a vendor-agnostic method to describe battery cycling protocols that we name 'battery cycler language' (BCL). This section continues to describe the publication data extraction process and corresponding structure and storage according to the FAIR principles.

**2.1 Semantic data treatment**

Semantic data treatment creates a metadata markup with standardised and resolvable vocabulary terms that can provide links to a shared conceptualization and data from other sources. This approach has already been applied with great success in the field of bioinformatics, and more generally as a foundation for the World Wide Web. Linked data is designed to be shared through a machine-readable extension of the World Wide Web called the Semantic Web. Our approach leverages similar semantic technologies to create battery linked data that is machine-readable and compliant with the FAIR data guidelines.

Ontologies are the foundation for semantic data markup. They provide a human- and machine-readable language that formalises knowledge using a common vocabulary. Although a foundational set of ontologies exists for describing simple resources, knowledge organisation, tabular data, or popular search engine queries, there has been a lack of comparable ontologies for describing batteries. The European Union project BIG-MAP [13] developed the Battery Interface Ontology (BattINFO) [14], [15] to address this need. BattINFO is a collection of relevant domain ontologies from the top-level Elementary Multiperspective Materials Ontology (EMMO) universe. It adheres to the World Wide Web Consortium (W3C) standards for ontologies and linked data, allowing data that has been marked-up with BattINFO terms to be integrated into the Semantic Web. Battery 2030+ provides support for the continued development and maintenance of BattINFO as a resource for the battery research community [16].

In this work, we implement the mapping between battery data and its associated ontology terms using JavaScript Object Notation for Linked Data (JSON-LD). JSON-LD is a serialisation format within the Resource Description Framework (RDF), the standard for expressing semantic data on the web. JSON-LD is particularly advantageous because it is human-readable and can be parsed into a machine-readable knowledge graph. We use terms from BattINFO to create linked data descriptions of battery cells and test procedures, while relying on established RDF vocabularies like Schema.org for describing general information like manufacturers, product IDs, names, etc. This approach provides two main benefits: it ensures that the data from this work is easily re-usable and interoperable, and it provides an opportunity to perform semantic queries using the powerful concept-based protocol and RDF query language SPARQL.

**Battery cycler language**

The experimental procedure of cycling batteries is very specific to the field and requires specific vocabulary. Therefore, the terminology for battery cycling cannot be based on existing standards. Furthermore, nomenclature differs between cycler brands including symbols and units. Thus, we use the terminology from BattINFO [18] to describe the cycling procedures (see GitHub repo at [17]). The cycling procedures can than be exported as JSON-LD to facilitate interoperability. This work aims to integrate all descriptors from cycler brands to make them compatible with the unified language proposed herein and facilitate interoperability.

Figure 2 shows the JSON file for a protocol named 'MinimalExample' written in the cycler language with a constant current charge until the upper cut off voltage. After the definition of relevant parameters such as cell capacity, the instructions for the cycler are given. In this example the protocol consists of only one sequence, i.e., a constant current charge at 1C that is terminated when the upper cut off voltage is reached. The resulting JSON file is then converted into a JSON-LD file in which further context and machine-readable definitions are added.

This minimal example in Figure 2 can be exported into PyBAMM [19], which will extract each step and command of the procedure and conduct a simulation accordingly.

In the process of making battery testing more automated, this work is the first step toward an interface between battery cyclers and researchers. In the long-term we envision that both the up-stream (before a measurement) and down-stream (after a measurement) steps are automated. After taking measurements, data can be analysed and interpreted immediately and independently, and metadata is stored automatically. This means that one does not need to rely on any specific battery cycler software or tailored software scripts to analyse the data. Therefore, this is the first important step towards interoperable test procedures which also facilitates accessibility and reusability.

```json
"name": "MinimalExample",
"parameters": {
        "Capacity": 2.5,
        "UpperCutoffVoltage": 4.2
},
"instructions": [{
        "sequence": [{
                "type": "ElectricCurrent",
                "value": 1,
                "unit": "CRate",
                "termination": [{
                        "type": "Voltage",
                        "value": "UpperCutoffVoltage",
                        "unit": "Volt"
}]}]}]
```

Figure 2: Minimal example of a standardised test procedure written in the battery cycler language (BCL) with a constant current charge at 1C until the upper cut-off voltage of 4.2 V is reached.

## 2.2 Data curation and repository construction

Structured data makes the information easier to find and to reuse, both for humans and machines. We now show how we create structured data from cell data sheets that not only include the most typical properties such as capacity or voltage limits but are much more exhaustive and machine-readable. The additional information is extracted with an automated full-text search of 80,000 publications stored in a paper database.

### Create structured data

Numerous cells are commercially available and many of them have been tested in labs. The cell data sheets provided by the manufacturers vary widely in format and content, and the relevant information cannot be extracted automatically. Thus, as the first step, these data sheets are translated into machine-readable JSON-LD files based on the BattINFO ontology. In addition, data from Teichert et al. [20] were used. Our collection is limited to cells that have a manufacturer data sheet with the basic cell information but no further data on ageing and temperature-dependent discharge curves. Presently, the repository contains 400 files, one for each cell type. These contain typical information such as nominal capacity, voltage and temperature limits as well as the reference to where each piece of information comes from. The data in these files is additionally linked and organised in a knowledge graph, as schematically depicted in Figure 3. With this, the data is machine-readable and visualised.

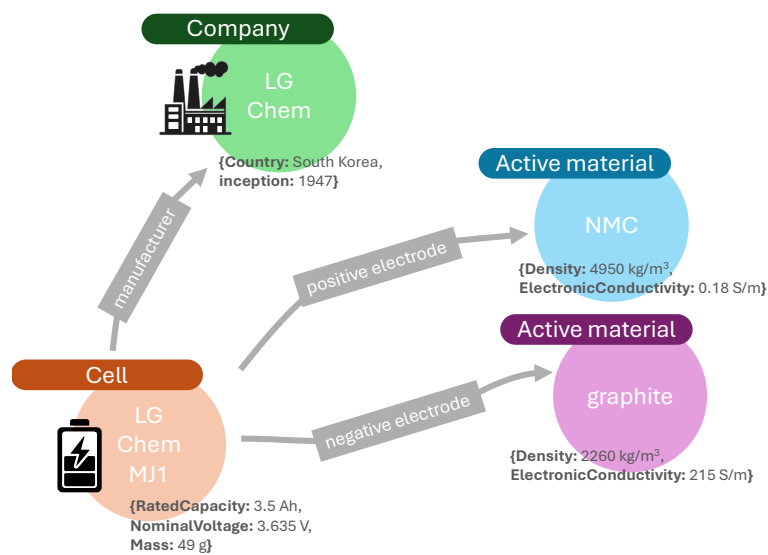

Figure 3: Battery cell data sheets contain useful information that can be parsed and organised in a knowledge graph. This example shows the connections between a specific LG Chem cell, its manufacturer, and its active materials (nickel manganese cobalt oxide (NMC) and graphite).

Furthermore, due to the interoperability of the process, additional user requested information can be appended to the structured data. We included, for example, the DOIs of the papers that have tested the individual cells (see Case Study 1 below). To find and access the data, a database containing most publications on batteries is created. This step is explained in the following section.

**Paper database**

As a key goal of this work is to enable an exhaustive full text search of battery information, a database is required containing all (or as many as possible) publications on batteries. For this work, papers needed to be published up until the end of 2023. Thus, a significantly larger paper database, aspiring to be exhaustive, was targeted than have been achievable in existing publications [5], [7], [21], [22].

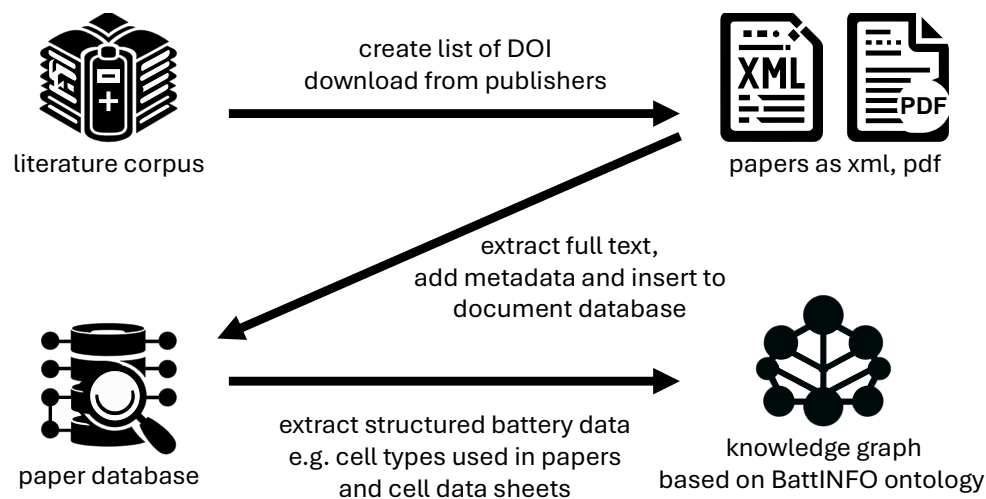

Figure 4: Information from articles was gathered and organised into a useful and searchable format by constructing a database from article text and metadata. Key structured data was then extracted and organised into a knowledge graph.

Figure 4 shows the flowchart of how the paper database is created. First, the literature available on Elsevier Scopus [1] and Semantic Scholar [23] was screened using the search term "Battery" to acquire a list of digital object identifiers (DOIs) of all publications relevant to battery research. All open-access articles were downloaded and the article text extracted from xml or pdf files into an elastic search database. To enable the inclusion of articles behind a paywall, a list of relevant publishers was selected. The papers were extracted from the list and assigned according to the publisher partial DOI. The chosen publishers are:

> Elsevier (10.1016, 10.3182)
> Wiley (10.1002, 10.1049)
> Springer/Nature (10.1007, 10.1038, 10.1134)
> ECS (10.1016)
> Taylor (10.1080)
> IOP (10.1088, 10.1149)
> IEEE (10.1109)
> MDPI (10.3390)
> Frontiers (10.3389)

These publishers were then either asked for permission for the automated download of their articles via an API key, or text was scraped according to text and data mining regulations for non-commercial research. Overall, this database includes more than 80,000 publications and is human- and machine-readable.

It must be mentioned that the dataset relies on underlying information from CrossRef, Semantic Scholar and Elsevier Scopus. Thus, it cannot contain publications not included in their data. For example, we noticed a gap in information around 2020. During this gap, not all publications covering batteries were indexed in Semantic Scholar. This gap was cross-referenced against data from Scopus, but we cannot guarantee that all missing papers have been identified. To reduce the chance of errors during downloading and processing, we conducted spot checks on a range of around 100 papers known to the authors.

Within this paper database a semantic search of the full text of all papers is possible, and this has two main advantages:

1. With semantic search the search term does not need to match exactly the term used in the papers, but related terms can also be found.
2. The search covers the entire text and not only title, abstract and keywords.

3. Results: Examples on how to use structured data

This work aims to advance battery research by implementing structured data methodologies that align with the FAIR principles—making data more Findable, Accessible, Interoperable, and Reusable. In the following we show several example case studies as results of the proposed methods. They demonstrate the potential of our approach and the value of structured data according to the FAIR principles for improving battery research.

- Case Study 1: Enhance findability through semantic searches and linking publications and cell types. With an exhaustive full text search of more than 80,000 articles structured data are created for batteries.
- Case Study 2: Enhance accessibility and reusability with large language models via retrieval-augmented generation for non-experts.
- Case Study 3: Enhance interoperability with the Battery Cycler Language.

**Case study 1: Complement structured data and find information in publications via full text search**

First, we want to extend the cell data sheets and add references that include information about a cell in the respective data sheet to improve findability of related information. As an example, with the semantic full text search in the paper database and the search term "LG Chem INR21700 MJ1", 44 papers with information on this cell could be identified. This information was verified manually and included in the JSON-LD file for the LG MJ1 (see file at https://github.com/phdechent/CellInfoRepository/blob/main/BatteryTypeJson/LG_Chem_INR18650_MJ1.json) To do so, it is required that the authors mention the specific cell type they tested somewhere in the paper. During the process we realised that this is seldom the case, and this hinders reproducibility checks, comparisons and transfer of findings. Thus, the method is only as good as the underlying information. The collected data can be found on https://github.com/phdechent/CellInfoRepository and is also presented on https://battdat.org.

Similarly, it is possible to search the full text of all publications for information and test data for cells from one specific manufacturer.

**Case study 2: Retrieve information stored in graph database**

One application of structured data is in combination with a large language model to enable non-experts to query the data and receive human-friendly responses which has the potential to make the information much more accessible. For simple queries, a search term can be used for the full text search in the database. For more complex queries, however, a large language model (LLM) is a useful tool to translate search terms from natural language into specialised search query strings for the databases. For this purpose, we used retrieval-augmented generation (RAG)—a combination of a LLM and the structured data (cell data sheet) as external information (non-parametric memory), to achieve precise results for search queries and to prevent the LLM from hallucinating [10]. Our approach is similar to Buehler [24] using an ontological knowledge graph as non-parametric memory. The relatedness of the stored data in the knowledge graph database provides advantages over regular retrieval augmentation approaches and improves LLM performance [24]. As the interaction with knowledge graphs is non-trivial and requires highly specific domain knowledge in formulating graph requests, this approach makes the knowledge graph database accessible for many more researchers and therefore increases its relevance.

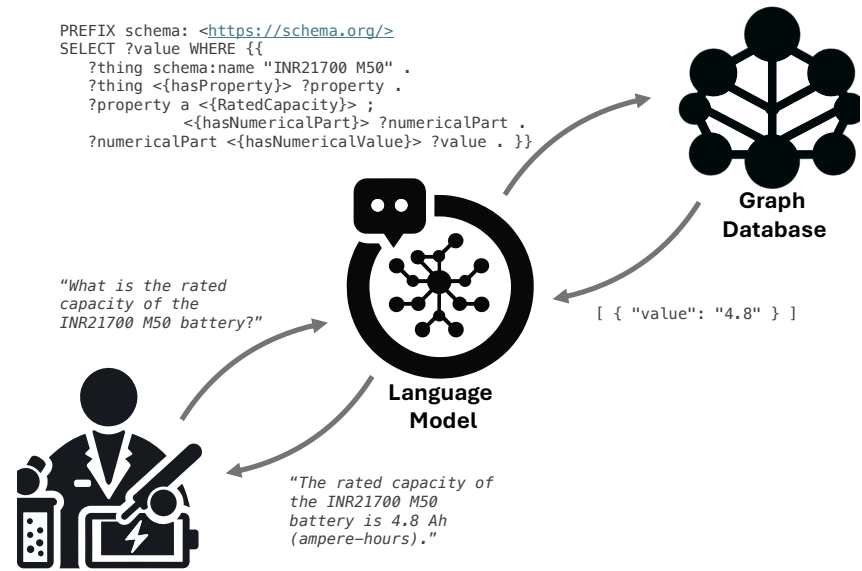

Figure 5: Battery-related example of a simple query being translated into SPARQL by the large language model and the answer being translated back into readable prose. Language models enable non-expert users to query complex structured databases and receive answers in human-readable form.

The schematic of this approach is shown in Figure 5 and other examples are provided in the supplementary information. A researcher asks the LLM a question in natural language about the capacity of a specific cell. The LLM uses its general knowledge and preprompting to "understand" the query and to translate it into a specific search query (English into SPARQL) for the knowledge graph database that acts as external information. Receiving the search query, the database returns an answer that is translated by the language model in natural language. The RAG can answer more complex questions than this simple example and there are additional examples given in the supplementary information.

To evaluate the performance of RAG on battery data, a set of 314 questions (see RAG example git) was formulated for several different LLMs, supplemented by the data in the knowledge graph database and LLMs (OpenAI gpt-4o-only, o1-only) without the additional graph knowledge. Figure 6 (a) shows the proportion of correct answers to overall questions. Comparing the results for gpt-4o and o1, the RAG performed significantly better than the LLM without additional information. However, the overall results would need to be improved before it becomes a valuable and reliable tool. The number one issue was the LLM formulating erroneous, non-compliant queries to the knowledge graph, thus leading to no information returned. There are several methods to improve the quality of query generation of non-specialized LLMs, for example with additional pre-prompting or low-rank adaptations (an inexpensive technique to fine tune large language models) [25].

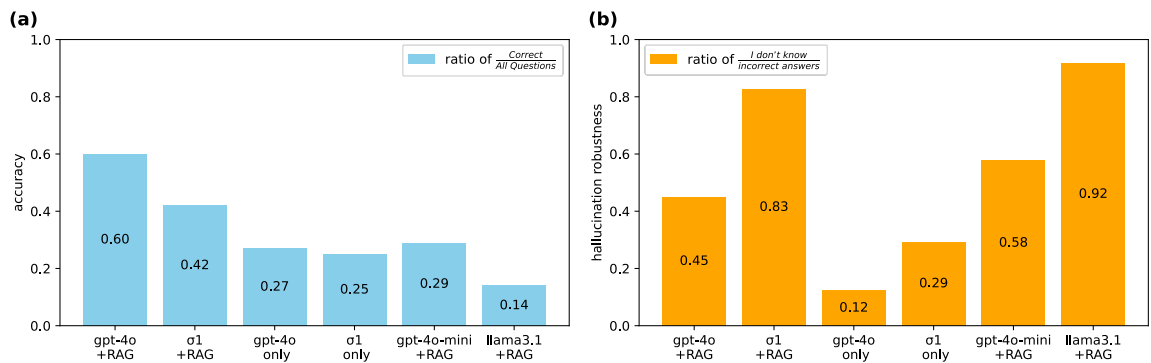

Figure 6: Adding structured data alongside a language model improves accuracy in response to battery queries, but overall results need further improvement. A set of 314 questions about batteries was evaluated with different LLMs in a retrieval augment generation setting with the battery graph data, as well as gpt-4o without the graph database; (a) shows the proportion of correct answers over all questions asked and (b) the hallucination robustness as the ratio of identifiable non answers (e.g. the model returns "I don't know the answer") to all non-correct answers.

Figure 6 (b) shows the robustness against hallucinations with the ratio of "I don't know" answers to incorrect answers. There is a clear trend visible: the trustworthiness through more robustness against hallucination is greatly improved with the RAG, since the LLM more often states it does not have the answer when it genuinely doesn't, instead of inventing an answer. OpenAI gpt-4o without the RAG structure gives the right answer in only 27% of cases and it also hallucinates "plausible" responses in 88% of non-correct answers. The knowledge in the LLM-only scenario comes from the model parameters during the training, as well as from the input. For example, "What is the capacity of the A123 23 Ah cell?" can be answered from the cell name itself, or the diameter of an 18650 cell be rightly assumed to be 18 mm. Overall this kind of LLM-only behaviour cannot be regarded as useful since it lacks trustworthiness. Even though reasoning models like OpenAI o1 outperform many models on topics like maths and coding, in our study they performed worse on the RAG task, similar to [26], while also being much more computational intensive and slower to respond.

Although the results leave room for improvement, RAG outperformed the LLM alone. These overall improved results show the power of RAG in making information findable and accessible.

**Case Study 3: Battery cycler language**

The battery cycler language can, for example, be applied to the cycling conditions in a cell data sheet that specify lifetime. For the LG MJ1, it states a cycle life of 400 cycles and specifies the conditions under which this was tested. The battery cycler language can capture nuances in the definitions. Figure 7 shows the extracted cycling conditions under which the battery should last 400 cycles before fading below 80% of the rated capacity in a C/5 capacity test. It is important to capture this information to make informed decisions while choosing a certain battery, since cycle life is often not stated on the maximum charge/discharge rates and sometimes conditions are also limited in cycling depth making a certain cell unsuitable for some applications.

```json
   "name": "Cyclelifecondition",
   "subjectOf": "LG Chem INR18650 MJ1", "id": "https://www.wikidata.org/wiki/Q120766894",
   "citation": "Specification%20INR18650MJ1%2022.08.2014.pdf",
     "parameters": {
         "Capacity": 3.4,
         "LowerCutoffVoltage": 2.5,
         "UpperCutoffVoltage": 4.2
     },
     "instructions": [{
         "sequence": [
             {"type": "ElectricCurrent", "value": 1.500, "unit": "Ampere",
                 "termination": [{"type": "Voltage", "value": "UpperCutoffVoltage", "unit": "Volt"}]
             },
             {"type": "Voltage", "value": 4.2, "unit": "Volt",
                 "termination":[{"type": "ElectricCurrent", "unit": "Ampere", "value": 0.100}]
             },
             {"type": "Rest", "value": 600, "unit": "Second"
             },
             {"type": "ElectricCurrent", "value": -4.000, "unit": "Ampere",
                 "termination": [{"type": "Voltage", "value": "LowerCutoffVoltage", "unit": "Volt"}]
             },
             {"type": "Rest", "value": 600, "unit": "Second"
             }],
         "name": "HighDrainrateChargeDischargecondition",
         "repeat": 400
     },
     {"sequence": [
             {"type": "ElectricCurrent", "value": 0.5, "unit": "CRate",
                 "termination": [{ "type": "Voltage", "value": "UpperCutoffVoltage", "unit": "Volt"}]
             },
             {"type": "Voltage", "value": 4.2, "unit": "Volt",
                 "termination": [{"type": "ElectricCurrent", "unit": "Ampere", "value": 0.050}]
             },
             {"type": "ElectricCurrent", "value": -0.2, "unit": "CRate",
                 "termination": [{"type": "Voltage", "value": "LowerCutoffVoltage", "unit": "Volt"}]
             }],
         "name": "cycle_401_reference_test",
         "repeat": 1}]
```

Figure 7: Codifying the test protocol used for life testing yields insights into whether a cell is suitable for an application or not. The battery cycler language gives a manufacturer-independent description of cycling conditions here from the datasheet of the LG Chem MJ1 cell. It specifies how the cell should be cycled, the discharge depth and currents for example, as well as a definition of a reference test in cycle 401 to determine that the capacity has not dropped below 80% of the rated capacity.

The next step is to make this language [17] interoperable with different cycler brands so that such procedures can be directly applied to cells under test.

4. Conclusion

This work has demonstrated a proof of concept in advancing battery research through structured data methodologies. We have successfully (i) developed a paper database containing almost all publications on batteries, enabling exhaustive semantic full-text searches; (ii) utilized structured data from cell data sheets within a knowledge graph; and (iii) paired large language models (LLMs) with structured battery data to facilitate advanced queries by non-experts. These efforts illustrate how battery researchers can benefit from structured data that adhere to the FAIR principles—making data more Findable, Accessible, Interoperable, and Reusable.

Our approach aligns with the FAIR principles and helps the community to adhere to these principles. By implementing structured data methodologies, we enhance findability through semantic searches and linking publications and cell types (Case Study 1), improve accessibility and reusability via retrieval-augmented generation for non-experts (Case Study 2), and promote interoperability with the Battery Cycler Language (Case Study 3). Thus, this work propels the battery field toward achieving FAIR data standards and enables more collaborative research efforts, contributes to more efficient knowledge discovery and innovation and accelerates research in the field.

However, this is just the beginning, and some limitations and challenges remain. The publications are limited to English literature, and downloading from publishers is not always straightforward. Copyright restrictions prevent us from making the paper database open source, leaving only the cell database freely accessible [27]. Additionally, these resources require continuous updates to keep pace with the high volume of publications and newly developed lithium-ion batteries. Recognizing this as a work in progress, we invite the battery community to participate—by suggesting terms and alterations in the ontology or by contributing to the cell database. Creating such structured data demands initial effort and ongoing maintenance, but it can evolve into a collaborative endeavour that benefits the entire community [28].

Looking ahead, the integration of LLMs with structured battery data opens intriguing possibilities. As LLMs become more sophisticated, they can make complex, structured datasets more accessible to a broader range of researchers, including those without specialized expertise. Advancements in retrieval-augmented generation techniques will enhance the accuracy and reliability of Large Language Model responses, reducing issues like erroneous queries and hallucinations. By refining natural language interfaces and incorporating advanced fine-tuning methods, future LLMs will offer more precise and trustworthy interactions with structured battery data.

**Acknowledgments**

P. Dechent and S. Paarmann are funded by the Deutsche Forschungsgemeinschaft (DFG, German Research Foundation) – 506629963; 511349305

E. Barbers, S. Clark, D. Howey and B. Planden thank Horizon Europe for funding for Digital Solutions for Accelerated Battery Testing (Digibatt) 101103997, Innovative and Sustainable High Voltage Li-ion Cells for Next Generation (EV) Batteries (IntelLiGent) 101069765, BIG-MAP 957189 and Battery2030+ (101104022). Views and opinions expressed are however those of the author(s) only and do not necessarily reflect those of the European Union or CINEA. Neither the European Union nor the granting authority can be held responsible for them.

D. Howey and Brady Planden thank UK Research and Innovation (UKRI) for funding under the UK government's Horizon Europe funding Guarantee 10107050 and 10038031.

**ORCID:**
Philipp Dechent 0000-0003-3041-1436
Elias Barbers 0000-0002-5836-2210
Simon Clark 0000-0002-8758-6109
Susanne Lehner 0000-0001-7542-9686
Brady Planden 0000-0002-1082-9125
Masaki Adachi 0000-0003-2580-2280
David Howey 0000-0002-0620-3955
Sabine Paarmann 0000-0003-1629-6830

**CRediT Author Contributions Statement:**
P.D.: conceptualisation, methodology, software, data curation, writing – original draft, visualisation.
E.B.: methodology, software, data curation.
S.C.: conceptualisation, methodology, validation, software, data curation, writing – original draft.
S.L.: validation, resources, writing– review and editing.
B.P.: methodology, software, writing– review and editing.

M.A.: writing– review and editing.
D.H.: methodology, supervision, writing – original draft.
S.P.: conceptualisation, methodology, writing – original draft.

**Disclaimer on the use of AI writing aids:**

During the preparation of the manuscript, the authors used GPT4-o1/OpenAI to improve readability and language. After using this tool, the authors reviewed and edited the content as needed and take full responsibility for the content of the publication.

**Data availability:**

There are two Git repositories with the cell data in JSON-LD format as well as example use cases https://github.com/phdechent/CellInfoRepository [27] and the battery cycling language with example cycling procedures available https://github.com/phdechent/BCL [17]. (The source code will be moved to an institutional repository before publication)

**Supplementary Information:**

Additional examples for Retrieval augmented generation questions:

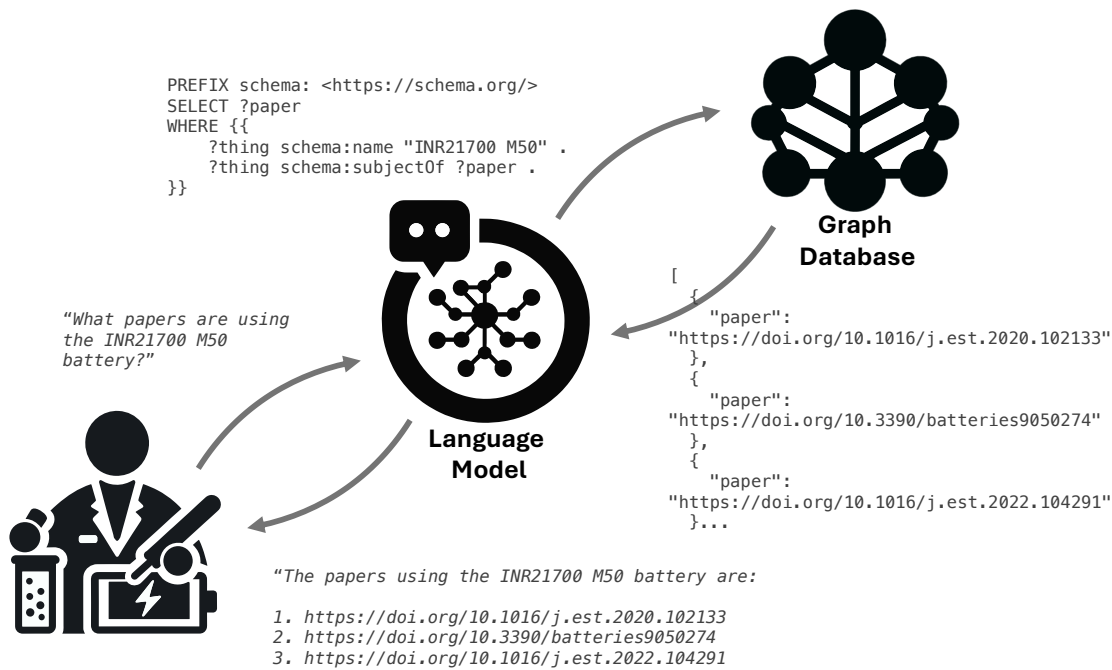

Figure S1: Example query for Retrieval augmented generation on the questions of which papers cover the LG M50 cell.

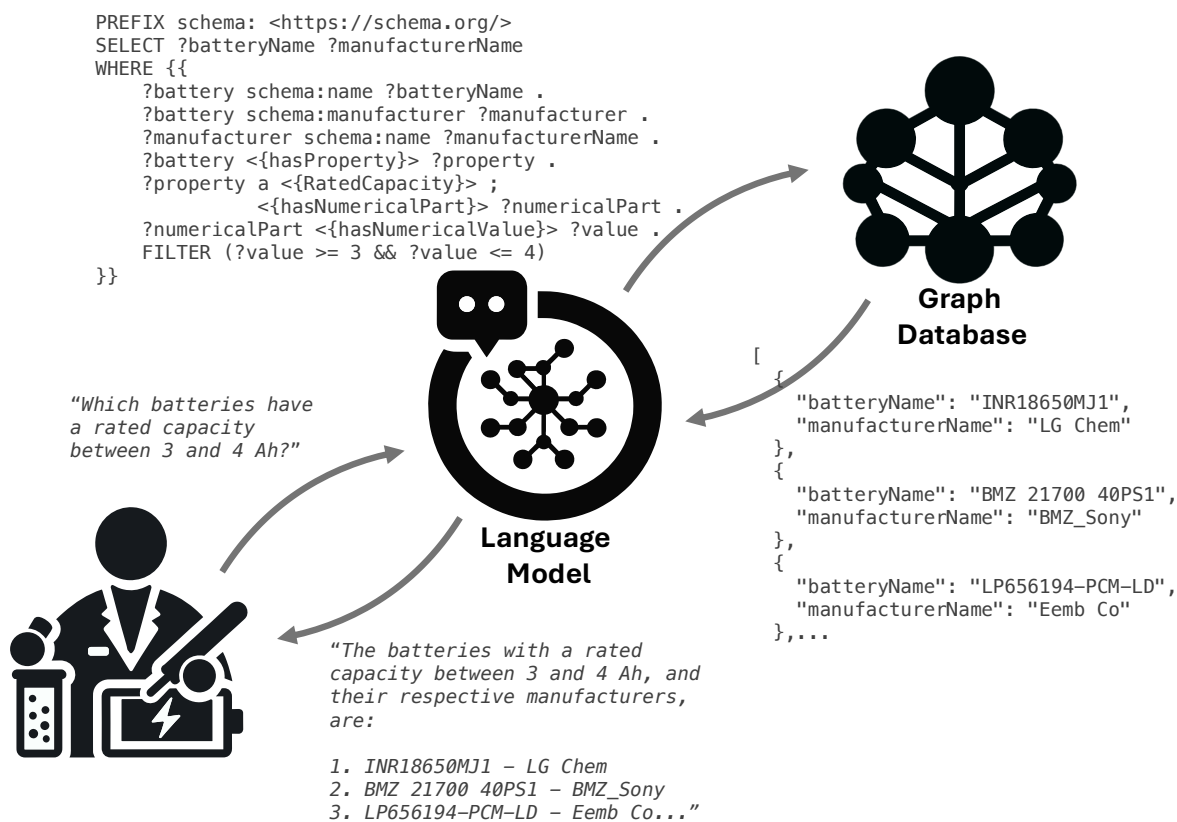

Figure S2: Example query for Retrieval augmented generation on the questions of which cell have a capacity of 3 to 4 Ah.